\documentclass[twocolumn,showpacs,preprintnumbers,amsmath,amssymb]{revtex4}

\usepackage{graphicx}
\usepackage{dcolumn}
\usepackage{bm}
\usepackage{amsmath}
\usepackage{url}
\usepackage{color}

\begin{document}

\title{Time of Flight and Supernova Progenitor Effects on the Neutrino Halo}

\author{John F. Cherry$^{1}$}
\author{George M. Fuller$^{2}$}
\author{Shunsaku Horiuchi$^{3}$}
\author{Kei Kotake$^{4}$}
\author{Tomoya Takiwaki$^{5}$}
\author{Tobias Fischer$^{6}$}

\affiliation{$^{1}$University of South Dakota, Vermillion, South Dakota 57069, USA}
\affiliation{$^{2}$Department of Physics, University of California, San Diego, La Jolla, California 92093, USA}
\affiliation{$^{3}$Center for Neutrino Physics, Department of Physics, Virginia Tech, Blacksburg, VA 24061, USA}
\affiliation{$^{4}$Department of Applied Physics \& Research Institute of Stellar Explosive Phenomena (REISEP), Fukuoka University, Fukuoka 814-0180, Japan}
\affiliation{$^{5}$National Astronomical Observatory of Japan, Mitaka, Tokyo 181-8588, Japan}
\affiliation{$^{6}$Institute of Theoretical Physics, University of Wroclaw, Pl. M. Borna 9, 50-204 Wroclaw, Poland}

\date{December 26, 2019}

\begin{abstract}
We argue that the neutrino halo, a population of neutrinos that have undergone direction-changing scattering in the stellar envelope of a core-collapse supernova (CCSNe), is sensitive to neutrino emission history through time of flight.  We show that the constant time approximation, commonly used in calculating the neutrino halo, does not capture the spatiotemporal evolution of the halo neutrino population and that correcting for time of flight can produce conditions which may trigger fast neutrino flavor conversion.  We also find that there exists a window of time early in all CCSNe where the neutrino halo population is sufficiently small that it may be negligible. This suggests that collective neutrino oscillation calculations which neglect the 
halo may be well founded at sufficiently early times.
\end{abstract}
\preprint{}

\maketitle

\section{Introduction}

We show novel features of the neutrino halo~\cite{Cherry:2012b} that may impact cutting edge research on neutrino signals and nucleo-synthesis in compact objects and core-collapse supernovae (CCSNe). During the collapse and subsequent supernova explosion of a massive progenitor star, some $\sim 3\times 10^{53}\,\rm erg$ of energy is released as neutrino radiation, a tiny fraction of which undergoes wide angle scattering to form a halo of neutrinos on trajectories which do not originate on the proto-neutron star (PNS) at the core.  

Wide angle neutrino scattering is also responsible for electron lepton number (ELN) crossings, where the net lepton number of neutrinos emitted changes sign across adjacent trajectories.  ELN crossing is understood to be key to the phenomenon of fast flavor conversion (FFC)~\cite{Sawyer:2005yg,Sawyer:2009aa,Dasgupta_2015,Sawyer_2016,Chakraborty:2016a,Chakraborty:2016b,Izaguirre_2017,Dasgupta_2017,Dasgupta:2018aa,Dasgupta_2018,Airen_2018,Abbar:2018a,ABBAR2019545,Capozzi:2019aa,Capozzi_2019,Yi:2019aa,Abbar:2019aa,Azari:2019aa,Shalgar_2019,Morinaga:2019aa,Nagakura_2019,Johns:2019a,Shalgar:2019bb} which are of great theoretical interest due to the potential of FFC to occur near the PNS surface, altering neutrino energy deposition rates.  Initial studies of the effects of wide angle neutrino scattering~\cite{Cherry:2013a,Dasgupta_2017,Izaguirre_2017,Cirigliano:2018a,Capozzi:2019aa,ABBAR2019545,Zaizen:2019aa,Shalgar:2019bb} have shown that halo neutrinos and ELN crossings have the potential to alter the flavor evolution of the entire emitted neutrino spectrum through $\nu-\nu$ coherent forward scattering and $\nu-\bar\nu$ pairwise correlation, which in a typical CCSNe may cause all neutrinos near the core to have quantum-correlated states.

\begin{figure}[h]
\centering
\includegraphics[scale=.28]{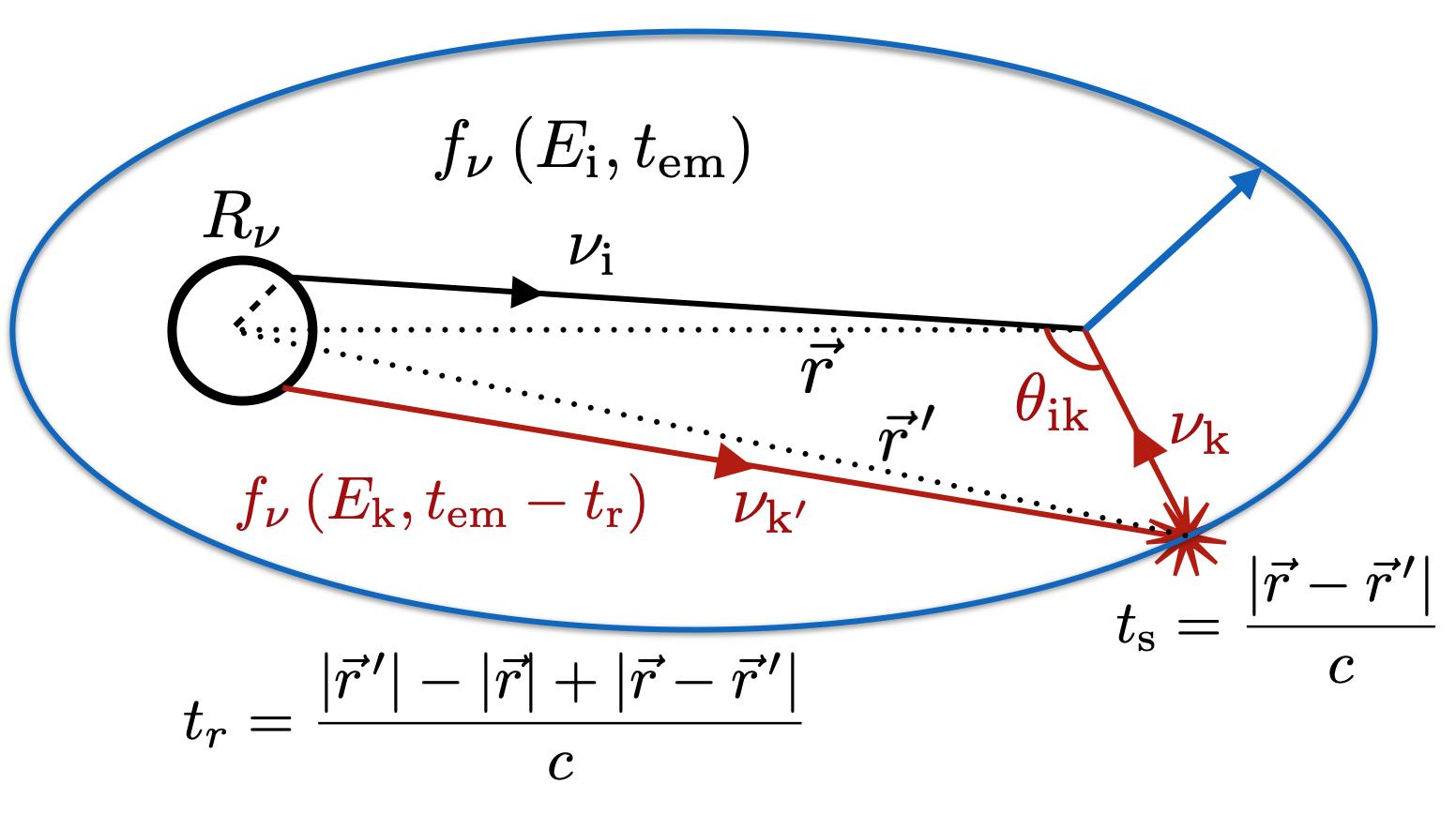}
\caption{Schematic neutrino halo temporal structure. 
{Halo neutrinos ($\nu_{\rm k}$, shown in red) arrive at the same location as radially emitted non-halo neutrinos ($\nu_{\rm i}$, shown in black) via a wide angle scattering at $\vec {r^\prime}$ and hence there is a ToF effect. It is illustrative to consider a simplifying picture where the neutrinos are emitted from a point source. Then, $t_r$ and $t_s$ are the time elapsed since the halo neutrino's emission and since the halo neutrino's scattering, respectively. The blue ellipsoid shows the region of the stellar envelope contributing halo neutrinos emitted from a constant look back time of $t_r$. The eccentricity of this ellipsoid illustrates the difference between CT and ToF treatments. The point source approximation is only for illustration purposes and not used for the calculations in this work. }
}
\label{Cartoon}
\vspace{-1em}
\end{figure}

Following the neutrino flavor evolution in the presence of scattering generally requires a solution of the quantum kinetic equations~\cite{stodolsky:1987QK,Sigl93,Strack:2005rr,Vlasenko:2013fja,Serreau:2014cfa,Richers:2019a}, which combine coherent flavor evolution with Boltzmann transport of neutrinos in a unified formalism.  Practical considerations have required simplifying assumptions in order to evaluate neutrino flavor evolution, such as the neutrino BULB model~\cite{Duan:2006b,Fogli:2007ys,Dasgupta:2008a} and what we refer to as the \lq\lq Constant Time\rq\rq\ approximation (CT) which appeals to the relativistic kinematics of neutrinos to replace the explicit time dependence of neutrino flavor evolution with the radial coordinate dependence, treating $t = \vert \vec r \vert/c$ as the single Affine parameter along the particle's world line.  The halo population of neutrinos, considered in the context of these simplifying assumptions, behaves similarly to a local, coherent coupling of neutrino flavor evolution histories.  

We seek to abandon the CT approximation and properly treat the composition of halo neutrinos and their contribution to the coherent evolution of neutrino flavor states emitted during CCSNe by accounting for finite time of flight (ToF) effects on neutrinos which are scattered into the halo population.  

\section{Time of Flight}

\subsection{Setup}

In Figure~\ref{Cartoon}, neutrinos are emitted in all directions from a neutrinosphere of radius $R_\nu$, with nearly all those emitted at time $t_{\rm em}$ arriving at radius $\vec r$ suffering only coherent forward scattering.  Halo neutrinos, in contrast, suffer a direction-changing scattering event at $\vec {r^\prime}$, and arrives at the same location via a trajectory that intersects $\vec r$ at a wide angle $\theta_{\rm ik}$.  

In analogy to the technique used in classical electrodynamics, we identify two \lq\lq retarded\rq\rq\ time scales relevant to enforcing causality between the radially emitted neutrino, $\nu_{\rm i}$, and the halo neutrino, $\nu_{\rm k}$: the time elapsed since $\nu_{\rm k}$'s scattering, $t_{\rm s}$, and the total time elapsed since $\nu_{\rm k}$'s emission, $t_{\rm r}$.  Within CCSNe, emitted neutrino luminosities, flavors, neutrino spectral energy distributions, entropy and mean nuclear masses, shock propagation, and turbulent mixing all can evolve on short $\sim 10-100\, \rm ms$ time scales.  {The retarded timescales are significant so long as either of two conditions are met: that the timescale for evolution of neutrino emission from the PNS at the time of emission, $t_{\rm em}$, is short compared to $t_{\rm r}$, or that the timescale for the evolution of the density and composition of the envelope of CCSNe at $\vec {r^\prime}$ is short compared to $t_{\rm s}$.}  {If either of these two cases obtain, the CT approximation will lead to a-causal calculations of the distribution of halo neutrinos arriving at $\vec r$.} 

{It is illustrative to imagine the PNS to be a point source in Figure~\ref{Cartoon}, so that the region of the stellar envelope which is contributing to the halo neutrino population which was emitted from the core at time $t_{\rm em} - t_{\rm r}$ is given by the ellipsoid of revolution with the PNS as one focus and the position $\vec r$ as the second focus.  The semi-major axis of this ellipsoid is then $a = c t_{\rm r} + \vert \vec r \vert$ with eccentricity $e = \vert \vec r \vert / ( ct_{\rm r} + \vert \vec r \vert )$.  As the eccentricity of this ellipsoid increases, the discrepancy between CT and ToF calculations of each shell's contribution to local neutrino number densities will grow.}

The ToF corrected expression for the number density of neutrinos in the halo (omitting factors which have no temporal dependence) is,{
\begin{eqnarray}
& n_{\nu}\left(\vec r,\theta_{\rm ik}, t \right) \propto 
 \int dE_{\rm k} d^3 \vec r^\prime dt_{\rm s} f_{\nu}\left( E_{\rm k},t_{\rm em} - t_{\rm s}\right) \rho\left( \vec r^\prime,t_{\rm em} - t_{\rm s}\right) &
\nonumber
\\
                 &  \times \mathcal M \left[ E_{\rm k}, \vec r - \vec r^\prime, A\left( \vec r^\prime,t_{\rm em} - t_{\rm s}\right), Z\left( \vec r^\prime,t_{\rm em} - t_{\rm s}\right) \right], &
\label{noftheta}
\end{eqnarray}}
where $E_{\rm k}$ is the neutrino energy, $f_{\nu}\left( E_{\rm k},t\right)$ is the emitted neutrino spectral energy distribution, the envelope matter density is $\rho\left( \vec r^\prime,t\right)$, the nuclear composition is given by the average atomic number $ Z\left( \vec r^\prime,t\right)$ and average nucleon number $A\left( \vec r^\prime,t\right)$, and $\mathcal M$ is the scattering kernel.  {Note that Equation~\ref{noftheta} is subject to the equation of constraint, $ct_{\rm s} = \vert \vec r -\vec r^\prime \vert$, which couples the integration variables $\vec r^\prime$ and $t_{\rm s}$. Further, the LHS time coordinate is related to the emission time such that $t = t_{\rm em} + \vert \vec r \vert/c$, with neutrinos emitted at $t > t_{\rm em}$ making no contribution to $n_{\nu}\left(\vec r,\theta_{\rm ik}, t \right)$.}  Previous work characterizing the effects of the halo neutrinos have treated Equation~\ref{noftheta} using the CT approximation{, taking $t_{\rm em} = t_{\rm post\ bounce} - \vert \vec r \vert /c$ and $t_{\rm s} = t_{\rm r} = 0$  identically~\cite{Cherry:2012b,Sarikas:2012vb,Cherry:2013a,Cirigliano:2018a,Zaizen:2019aa}}.  Similarly, prior work considering FFC~\cite{Dasgupta_2017,Dasgupta:2018aa,Dasgupta_2018,Airen_2018,Abbar:2018a,ABBAR2019545,Capozzi:2019aa,Capozzi_2019,Yi:2019aa,Abbar:2019aa,Azari:2019aa,Shalgar_2019,Johns:2019a,Shalgar:2019bb}  made use of 1D or ray-by-ray Boltzmann neutrino transport schemes which include neutrino propagation time delays only in 1D (along the axis of the ray).  This approach explicitly omits any ToF correction for neutrino transport {\it ray-to-ray}, implicitly employing the CT approximation at first order in the number of direction changing scatterings to compute neutrino distributions within each ray.  {At time of writing, the exceptions to this rule are the recent works of~\cite{Morinaga:2019aa} and ~\cite{Nagakura_2019} who have successfully implemented time-dependent multi-dimensional Boltzmann neutrino transport in their CCSNe simulations.  This allows~\cite{Morinaga:2019aa,Nagakura_2019} to properly resolve the ToF effects on direction changing neutrino transport.}

{
We calculate our ToF corrected neutrino number densities using the discrete time evolution of several CCSNe progenitor models~\cite{Woosley_2002,Kitaura06,Umeda_2008,Fischer_2010,Heger:2010a,Fischer_2012,Mart_nez_Pinedo_2012,Fischer_2016,Fischer_2018,Zaizen:2019aa}, which have been performed previously using 1D or ray-by-ray for neutrino transport.  These simulations are mapped into a discrete 3+1 dimension space-time grid, where the neutral current direction changing scattering component of the neutrino interactions with the matter in the envelope are computed in detail.  We do not assume a single neutrinosphere emitting surface, but instead calculate the integrated optical depth to all scattering processes, $\tau$, for each neutrino energy and species uniquely.  Individually by each species, zones with $\tau \geq 1$ are treated as emitting neutrinos isotropically, and zones with $\tau < 1$ are treated as \lq\lq decoupled\rq\rq, with neutrino transport taken to be radial directed with a small portion of neutrinos undergoing neutral current direction changing scattering.  This allows us to post-hoc recover the energy and species dependent limb-darkening of the neutrino emission region in the core, as well as the wide angle scattered population of the envelope.  It is still useful to define an effective neutrinosphere radius, $\langle R_\nu \rangle$, as the population averaged radius for which $\tau =1$,
\begin{equation}
    \langle R_\nu \rangle = \frac{\sum_{s = \nu,\bar\nu}\sum_{\alpha = \rm e,x} \int r_{\tau=1}\left(E,s,\alpha\right)E^2 f\left( E,s,\alpha \right) dE}
    {\sum_{s = \nu,\bar\nu}\sum_{\alpha = \rm e,x} \int E^2 f\left( E,s,\alpha \right) dE} \, ,
\label{Rnu}
\end{equation}
where $r_{\tau=1}\left(E,s,\alpha\right)$ is the radius for which each energy and species passes the threshold for integrated optical depth less than unity.}

\subsection{Collective neutrino oscillation}

{From our spatio-temporal map of direction changing neutrino scattering we have sufficient information to discretely evaluate Equation 1, using appropriate geometric factors, from which we can then} directly construct the neutrino-neutrino Hamiltonian, ${H}_{\nu\nu}$, which couples the flavor histories for neutrinos on intersecting trajectories~\cite{Sigl93,Qian95,Duan:2010fr}.  As shown in Figure~\ref{Cartoon}, a neutrino $\nu_{\rm i}$ leaving the core will experience a potential given by a sum over neutrinos located at the same point as neutrino $\nu_{\rm i}$:{
\begin{eqnarray}
& {H}_{\nu\nu}
 =  \sqrt{2}\,G_{\rm F}\,\int \left( 1-\cos{\theta_{\rm ik}}\right)  \times & 
 \\
                 &   [ \left( n_{\nu,\rm e}\left( \theta_{\rm ik}\right)  - n_{\nu,\rm x}\left( \theta_{\rm ik}\right) \right) -
                  \left( n_{\bar\nu,\rm e}\left( \theta_{\rm ik}\right)  - n_{\bar\nu,\rm x}\left( \theta_{\rm ik}\right)\right) ] d \cos\theta_{\rm ik}, &
\nonumber
 \label{Hnunu}
\end{eqnarray}}
where $\theta_{\rm ik}$ is the angle of intersection between $\nu_{\rm i}$ and neutrino or antineutrino $\nu_{\rm k}/\bar{\nu}_{\rm k}$.  {Here, $n_{\nu,\rm e/x}$ is the local number density of neutrinos in each flavor state (with the angular integration likewise evaluated locally)} and the $1-\cos\theta_{\rm{ik}}$ factor disfavors small intersection angles, thereby suppressing the potential contribution of the forward-scattered-only neutrinos~\cite{Fuller87,Fuller:1992eu}.  Halo neutrinos may have larger intersection angles as shown in Figure~\ref{Cartoon}, and therefore can contribute significantly to the flavor-changing potentials despite their small numbers.  {It should be noted that Equation 3 is the proper $\nu-\nu$ forward scattering Hamiltonian contribution in the limit that all neutrinos remain in their initial flavor states.  While it is not sufficient to use this construction generally, our intent in this work is to diagnose the presence of neutrino distributions that would likely result in collective neutrino oscillation or fast pair-wise flavor conversion if such effects had not already taken place.  As such, Equation 3 captures the initial conditions necessary to test for FFC.} 

\begin{figure}[h]
\centering
\includegraphics[scale=.43]{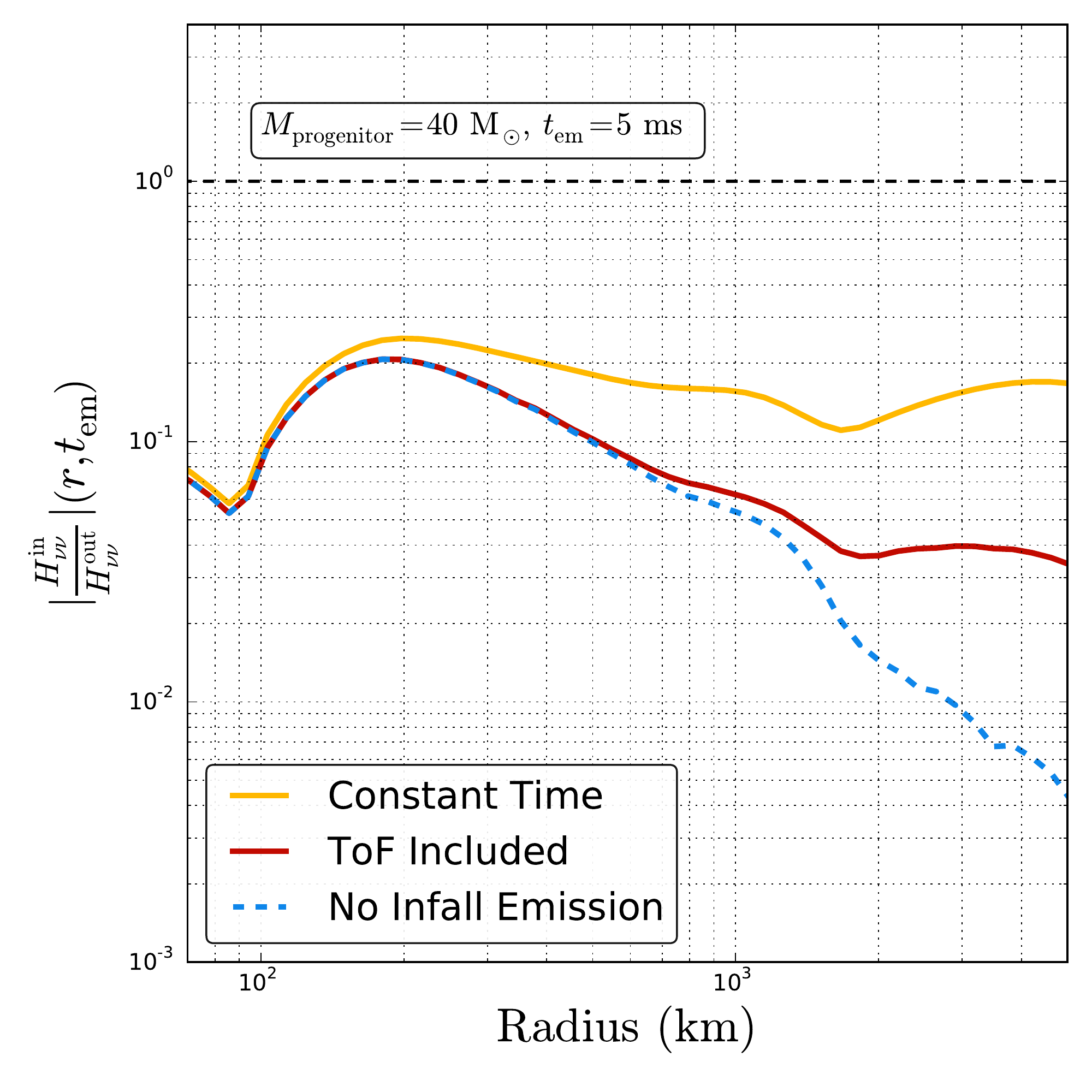}
\caption{ToF effects on the magnitude of the ratio of the inward directed neutrino Hamiltonian contribution to the outward directed neutrino Hamiltonian contribution {at a single time snapshot at $t_{\rm em} = 5$ ms for a $40\, M_\odot$ progenitor~\cite{Woosley_2002}. The reduction in the inward directed contribution due to ToF effect is substantial at this epoch. The ``No Infall Emission'' excludes the contribution from neutrino emission between the onset of gravitational instability and core-bounce.}}
\label{ToF}
\end{figure}

In the mean-field, coherent approximation, neutrino flavor evolution is governed by the equation $i {{\partial \vert\psi_{\nu,\rm i}\rangle  }/{\partial t  }}={H} \vert\psi_{\nu,\rm i}\rangle$ \cite{Halprin:1986cr}, where  $t$ is the Affine parameter along $\nu_{\rm i}$'s world line, and ${H} = {H}_{\rm V} + {H}_{\rm e} + {H}_{\nu\nu}$ is the appropriate neutrino propagation Hamiltonian, with vacuum and  matter components ${H}_{\rm V}$ and ${H}_{\rm e} $, respectively.   At any point within the envelope, $ H_{\nu\nu}$ can be split into two pieces: the contribution from outward directed neutrinos $ H^{\rm out}_{\nu\nu} = \left[  H_{\nu\nu} \right]_{0}^{\pi/2}$, and the contribution from inward directed neutrinos $ H^{\rm in}_{\nu\nu} = \left[  H_{\nu\nu} \right]_{\pi/2}^{\pi}$, with $ H_{\nu\nu} =  H^{\rm out}_{\nu\nu} +  H^{\rm in}_{\nu\nu}$.  We make this split because of the intrinsic asymmetry of neutrino emission in the CCSNe environment, with the radially directed neutrino emission from the PNS core providing the preponderance of neutrino number density in the envelope. 

The extent to which CT is a reasonable approximation can be quantified by comparing the ratio $\vert  H^{\rm in}_{\nu\nu} \vert / \vert  H^{\rm out}_{\nu\nu} \vert$ as a function of $ \left( r,t_{\rm em}\right)$ to the same quantity when calculating the halo neutrino population including the ToF correction in Equation~\ref{noftheta}.  Figure~\ref{ToF} shows the result of this comparison for a single time snapshot for neutrinos emitted $5\,\rm ms$ post core bounce for a CCSN simulation of a $40\, \rm M_\odot$ progenitor star~\cite{Woosley_2002}.  We can see that the CT approximation significantly overestimates the magnitude of $H^{\rm in}_{\nu\nu}$ at this early time.  Under the CT approximation, the entirety of the envelope is taken to be scattering neutrinos into the halo assuming the neutrino emission luminosity is $L_{\nu}( 5\,\rm ms )$, which is quite large early in the neutrino burst.  During the core infall epoch neutrino emission which populates the halo is considerably less luminous and lower energy.  From Figure~\ref{Cartoon} it can be seen that the volume of the stellar envelope which is illuminated by the neutrino burst at $t_{\rm em} = 5\,\rm ms$ is greatly reduced, bounded spatially and temporally by the requirement that $t_{\rm r} \leq 5\,\rm ms$.  

Figure~\ref{ToF} shows that the infall epoch neutrino emission from the core, emitted in the time between the onset of gravitational instability and core-bounce at nuclear densities ($\sim 0.3\, \rm s$), contributes to the halo neutrino population, accounting for an overall magnitude of $\sim {\rm (a\ few)}\,\%$ of the ratio $\vert  H^{\rm in}_{\nu\nu} \vert / \vert  H^{\rm out}_{\nu\nu} \vert$.  This raises the point that the ToF correction for the halo neutrino potential is coupled to the progenitor star and its subsequent CCSN evolution.  Progenitor properties such as pre-collapse mass, neutronization history, core-compactness, composition and density structure play a role in infall neutrino emission and pre-population of the halo neutrinos.  After core bounce the hydrodynamic evolution of the CCSN affects the emitted neutrino luminosity history through accretion and through halo neutrino scattering on nuclei/nucleons in the envelope. Shock propagation and turbulence influences these issues.

\begin{figure*}[htbp]
	\begin{minipage}{0.30\hsize}
	    \centering
    	\includegraphics[width=1.0\linewidth]{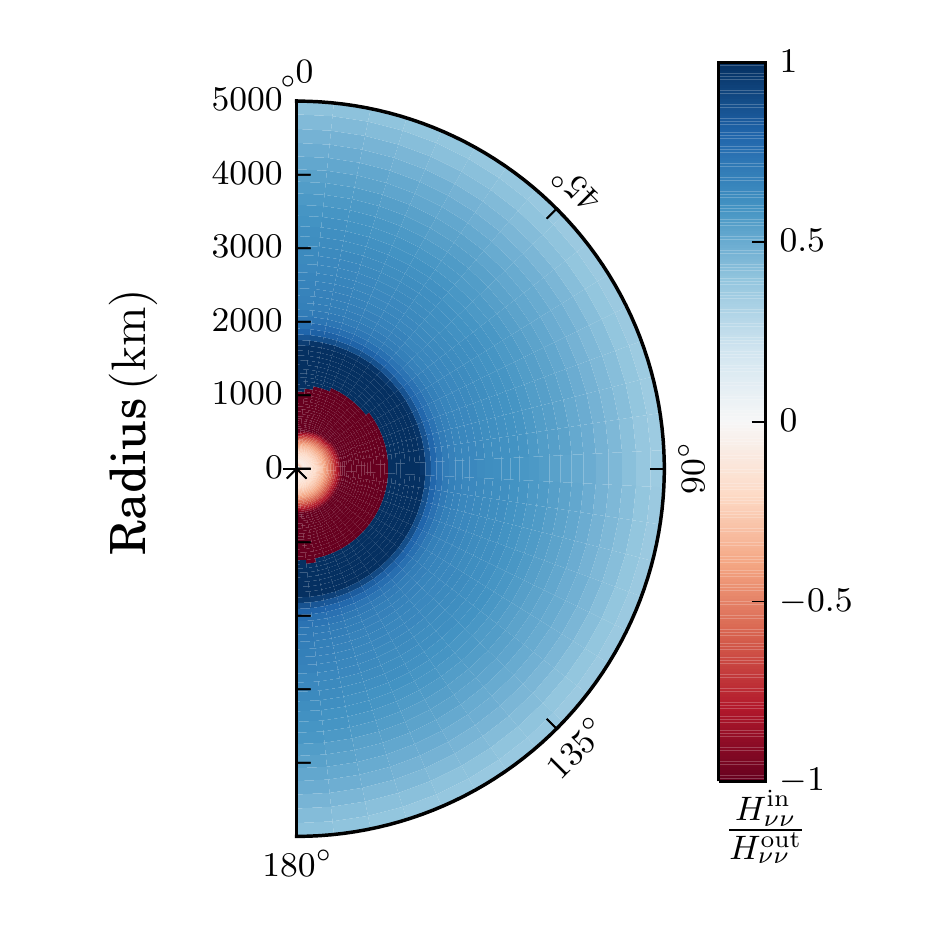}
	\end{minipage}
	\begin{minipage}{0.30\hsize}
	    \centering
    	\includegraphics[width=1.0\linewidth]{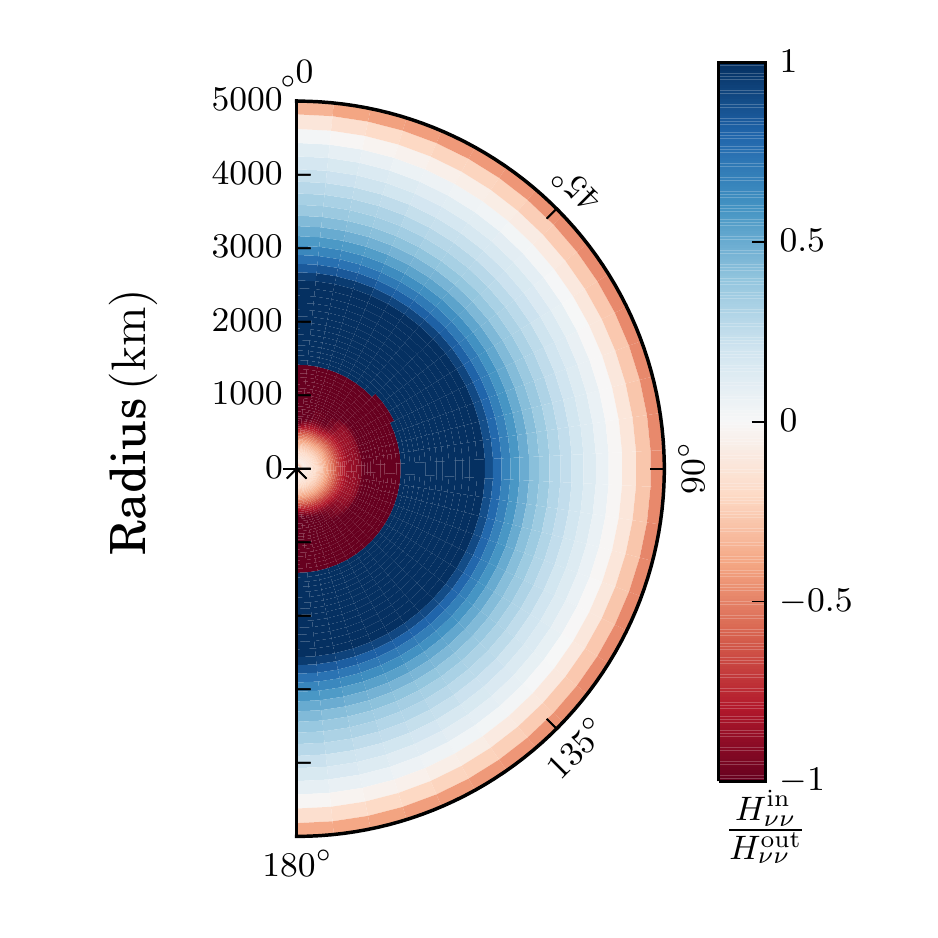}
	\end{minipage}
	\begin{minipage}{0.30\hsize}
	    \centering
    	\includegraphics[width=1.0\linewidth]{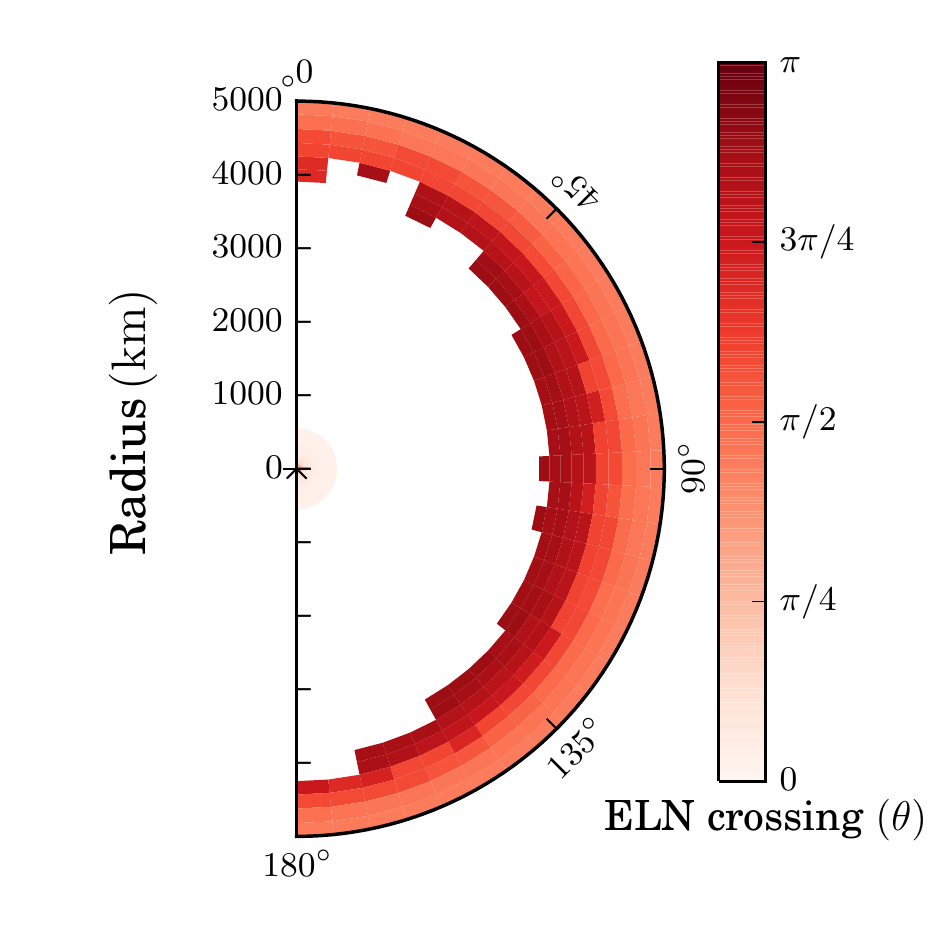}
	\end{minipage}

\caption{Left and Center panels: The relative sign change in $H^{\rm in}_{\nu\nu}/H^{\rm out}_{\nu\nu}$ as a function of position in the envelope for a $11.2\, M_\odot$ progenitor star {for a neutrino emitted $t_{\rm em} = 55\, \rm ms$ after core bounce.}  Left panel shows results for the CT approximation.  Center panel shows results for the ToF corrected halo.  Right panel: ToF corrected ELN crossing angle for $\nu_{e} - \bar\nu_{e}$ for the same $t_{\rm em} = 55\, \rm ms$ post core bounce snapshot.  The ELN crossing angle, $\theta$, is defined such that $0$ is radially outward and $\pi$ is radially inward.}
\label{SignFlip}
\end{figure*}

\subsection{Progenitor dependence}

The impact of TOF on the neutrino flavor transformation can be affected by the progenitor structure.  A generic effect of ToF at early times is to reduce the ratio $\vert  H^{\rm in}_{\nu\nu} \vert / \vert  H^{\rm out}_{\nu\nu} \vert$ by reducing the volume of brightly illuminated stellar envelope contributing to $H^{\rm in}_{\nu\nu}$ ahead of the ellipsoid focus at $\vec r$.  So long as $\vert  H^{\rm in}_{\nu\nu} \vert / \vert  H^{\rm out}_{\nu\nu} \vert$ is much less than unity, it may be reasonable to treat the flavor evolution of neutrinos using established techniques.  The ratio must inevitably rise as more of the stellar envelope is illuminated with high luminosity neutrino radiation.  To quantify the window of time in which ToF reduction of the magnitude of $H^{\rm in}_{\nu\nu}$ is significant in the calculation of neutrino flavor evolution, we define the timescale $t_{\mathcal O \left( 1\right)}$ to be the minimum time at which $\vert  H^{\rm in}_{\nu\nu} \vert / \vert  H^{\rm out}_{\nu\nu} \vert \geq 1$ for any radius $\vec r$ in the stellar envelope.  

Table~\ref{tab:tO1} shows the results of this calculation for a variety of progenitors of varying mass.  At the low end of the mass scale, $8.8\,\rm M_\odot$~\cite{Kitaura06,Sasaki:2019a} and $9.6\, \rm M_\odot$~\cite{Heger:2010a,Zaizen:2019aa}, we find that with ToF corrections the halo neutrino population is not sufficient to push $\vert  H^{\rm in}_{\nu\nu} \vert / \vert  H^{\rm out}_{\nu\nu} \vert \geq 1$ over the course of the simulation, and so these simulations potentially have $t_{\mathcal O \left( 1\right)} \rightarrow \infty$.  {This opens the possibility that conventional techniques may be used to calculate the neutrino flavor transformation for stars in this mass range.  We stress that it is by no stretch of the imagination {\it certain} that traditional techniques will be sufficient. Additional investigation into the safety of such techniques, such as those used in~\cite{Zaizen:2019aa}, may be fruitful.} 

\begin{table}[h!]
  \begin{center}
    \caption{Halo Growth Timescales for {select progenitors~\cite{Woosley_2002,Kitaura06,Umeda_2008,Fischer_2010,Heger:2010a,Fischer_2012,Mart_nez_Pinedo_2012,Fischer_2016,Fischer_2018,Zaizen:2019aa}}. Shown are the minimum times at which $\vert  H^{\rm in}_{\nu\nu} \vert / \vert  H^{\rm out}_{\nu\nu} \vert \geq 1$ for any radius in the stellar envelope, {taking the reference time $t=0\,\rm ms$ to be the core bounce time.} }
    \label{tab:tO1}
    \begin{tabular}{|l|c|c|c|c|c|c|c|} 
       \hline
         Mass ($M_\odot$) & 8.8 & 9.6 &10.8 &11.2 &15.0 &40.0 &50.0 \\
         \hline
	$t_{\mathcal O \left( 1\right)}\, \left( {\rm ms} \right)$ &   $> 335$& $> 230$&  $95$&  $53$&  $64$&  $60$&  $66$\\
        \hline
    \end{tabular}
  \end{center}
\end{table}

As progenitor mass increases, $t_{\mathcal O \left( 1\right)} = 95\, \rm ms$ for a $10.8\, \rm M_\odot$ progenitor~\cite{Woosley_2002,Fischer_2010} and rapidly converges to $t_{\mathcal O \left( 1\right)} \sim 60\, \rm ms$ for higher mass progenitors~\cite{Woosley_2002,Umeda_2008,Fischer_2012,Mart_nez_Pinedo_2012,Fischer_2016,Kotake_2018,Fischer_2018}.  The near independence of $t_{\mathcal O \left( 1\right)}$ on progenitor mass seems to come from a counter balance of competing effects.  The more massive progenitors tend to be more centrally condensed prior to collapse, which increases the baryon (and scattering target) density near the PNS and increases $\vert  H^{\rm in}_{\nu\nu} \vert$.  {This same increase in central condensation tends to increase the mass accretion rate and slow the contraction of the effective neutrinosphere radius, $\langle R_\nu \rangle $.  Likewise, rapid accretion on the PNS also tends to increase neutrino luminosities and temperatures, leading to a larger contribution to $\vert  H^{\rm in}_{\nu\nu} \vert$ through the energy dependence of the direction changing neutral current scattering.  It is not apparent why these effects should cancel, or that they do in fact cancel given the relative low number of SNe models considered here, but it is intriguing that the high mass progenitor models in Table~\ref{tab:tO1} show similar $t_{\mathcal O \left( 1\right)}$ timescales.}

\subsection{Time dependence}

Another feature of the ToF corrected neutrino halo population can be seen when comparing maps of the ratio $H^{\rm in}_{\nu\nu}  /   H^{\rm out}_{\nu\nu} $ (without absolute values) between ToF and CT calculations, 2D examples of which are shown in Figure~\ref{SignFlip} for the $t_{\rm em} = 55\,\rm ms$ snapshot of the $11.2\, M_\odot$ progenitor CCSN simulation.  We expect disagreement between CT and ToF to be larger early in the CCSN history because of ToF reduction in the illuminated volume of the envelope, and greater at large radius because of the increased eccentricity of the ellipsoid which populates the ToF corrected halo neutrino population.  Both of these trends can be seen in Figure~\ref{SignFlip}, which disagree by so much as the overall sign of $H^{\rm in}_{\nu\nu}\left( r \right)$ beyond $\sim 4500 \,\rm km$. { As a matter of practicality, the full computational domain of the CCSNe simulations we consider are extended out to a radius of $1.5\times 10^5\, \rm km$ using the density and composition of the original progenitor models where available or synthetically extended otherwise.  This allows a total time domain of $\sim 1\,\rm s$ over which we can self-consistently calculate the ToF effect on the halo population without loss of neutrino emission history.  The models we consider all have pre-core bounce collapse phases which are shorter than $300\,\rm ms$, so the results shown in Figures 2 and 3 and in Table 1 capture the full time evolution history of the neutrino emission of the core at all times considered.}

This sign change in $H^{\rm in}_{\nu\nu}\left( r \right)$ is a new phenomenon not found within the framework of the CT approximation.  Because of the energy dependence of neutrino direction changing scattering processes (e.g., the neutral current scattering cross section is $\propto G_F^2 \langle E_\nu^2\rangle$), it is possible under the CT approximation that $H^{\rm in}_{\nu\nu}$ has a relative sign difference compared to $H^{\rm out}_{\nu\nu}$ if,
\begin{align}
& Sign\sum_\alpha \left[ \frac{L_\alpha}{\langle E_\alpha\rangle}-\frac{L_{\bar\alpha}}{\langle E_{\bar\alpha}\rangle}   \right] \neq \nonumber \\
& Sign\sum_\alpha \left[ \frac{L_\alpha G_F^2\langle E^2_\alpha\rangle}{\langle E_\alpha\rangle}-\frac{L_{\bar\alpha} G_F^2\langle E^2_{\bar\alpha}\rangle }{\langle E_{\bar\alpha}\rangle}   \right]\sigma\, ,
\label{Sgnflp}
\end{align}
where $L_{\alpha/\bar\alpha}$ is the neutrino luminosity for each flavor and $\sigma$ is the integrated column density of scattering targets (which also scales with the composition of scattering targets through the mean nuclear mass squared dependence of coherent neutral current scattering).  Here the left hand side is the limiting case for the radially emitted net lepton number, and the right hand side is proportional to the limiting case for the wide angle scattered neutrino net lepton number.  Satisfaction of Equation~\ref{Sgnflp} is sufficient to guarantee the existence of an ELN crossing in the outward directed neutrinos.  However, because the halo neutrino energy spectrum is identical at all points under CT, once the ELN crossing point has been reached the relative sign of the net lepton numbers and Hamiltonian components will not be reversed.

Including the ToF correction adjusts the energy spectrum of the halo neutrinos to include integrated emission and scattering history over the entire envelope.  The corrected energy spectra-angular distribution at a given point can be dominated by neutrinos emitted in the remote past relative to when $t_{\rm em}$ lies within the explosion timescale.  Even though the neutrino emission conditions at $t_{\rm em}$ may satisfy Equation~(\ref{Sgnflp}), the integrated history of the halo neutrinos may not, overwhelming the nearby scattering contributions which are driving $H^{\rm in}_{\nu\nu}$ to have a sign change in some regions of the envelope.  This must necessarily be accompanied by an additional ELN crossing on inward directed trajectories, which can be seen on the right hand side of Figure~\ref{SignFlip}.

Note that the inward directed ELN crossing is a feature of the temporal structure of the ToF halo neutrino population.  The earliest portions of the post core-bounce neutrino emission are characterized by the deleptonization burst, where the shock emerges from the PNS, occasioning a significant neutrino luminosity increase over $\sim 20-30\, \rm ms$, with an electron capture-induced preponderance of $\nu_e$ over other species. Consequently, the halo neutrino population arising from this epoch is dominated by $\nu_e$.  The effect of ToF on the net lepton number in the halo is to dilute the contribution from recently emitted neutrinos, $t \lesssim t_{\rm em}$.  As a result we see in Figure~\ref{SignFlip} that although the emission satisfies Equation~\ref{Sgnflp} at {small radii, an ELN crossing develops on inward directed trajectories as the deleptonization burst neutrinos begin to dominate the halo neutrino population at large radius.  Thus the neutrino radiation field contains two distinct ELN crossing: the outward directed ELN crossing located at small radii and an inward directed ELN crossing at large radii.  It should be noted that the ToF induced inward directed ELN crossing which we report in this paper is distinct from other recently reported novel ELN crossing~\cite{Morinaga:2019aa,Nagakura_2019}, which are attributed to differential wide angle scattering from the heavy nuclei located near the surface of the accretion shock in Ref.~\cite{Morinaga:2019aa} and multidimensional fluid effects on $\nu/\bar\nu$ fluxes in Ref.~\cite{Nagakura_2019}.}

\section{Conclusions}

In this article we have demonstrated that time of flight effects for neutrinos emitted in CCSNe are large enough that they must be included in calculations which seek to model collective neutrino oscillations and fast flavor conversion.    We have argued that constant time treatments and ray-by-ray Boltzmann transport treatments are insufficient to capture the full temporal structure of the halo neutrino populations.  We have found that ToF effects produce a more complicated relationship between ingoing and outgoing neutrinos that can alter the relative sign of the coherent forward scattering Hamiltonians, $H^{\rm in}_{\nu\nu}$ vs.~$H^{\rm out}_{\nu\nu}$, and produce inward directed ELN crossings which arise as features of the integrated time evolution history of the CCSN neutrino emission and subsequent hydrodynamic evolution.  Further, we have discovered that CCSN progenitor effects work in favor of creating an environment where canonical techniques for calculating neutrino flavor evolution are viable at sufficiently early times.  A relatively stable window of time exists early in CCSNe where the wide angle scattered neutrinos may make a manageably small contribution to neutrino flavor oscillation and conversion, so that looking forward we will have a reliable \lq\lq stepping off point\rq\rq\ for theoretical models of CCSN neutrino signals.

\section{Acknowledgments} 

The work of GMF is supported by NSF grant No.\ PHY-1914242, by the NSF N3AS Hub Grant No.\ PHY-1630782, and Heising-Simons Foundation Grant No.\ 2017-228. The work of SH is supported by the US Department of Energy Office of Science under award number DE-SC0020262, and NSF Grant numbers AST-1908960 and PHY-1914409. KK acknowledges support by Grant-in-Aid for Scientific Research (JP17H01130) from the Japan Society for Promotion of Science (JSPS) and the Ministry of Education, 
Science and Culture of Japan (MEXT, Nos.\ JP17H06357 and JP17H06364), and by the research grant (Nos.\ 171042,177103) at Fukuoka University. Supernova simulations were performed at the Wroclaw Center for Scientific Computing and Networking (WCSS) in Wroclaw (Poland). TF acknowledges support from the Polish National Science Center (NCN) under grant number UMO-2016/23/B/ST2/00720. This work was supported by the COST Action CA16117 ``ChETEC''.

\bibliography{allref}

\end{document}